\documentclass[aps,prl,twocolumn,showpacs,groupedaddress]{revtex4}
\usepackage{graphicx}
\usepackage{dcolumn}
\usepackage{bm}
\usepackage{ae}
\usepackage{aecompl}
\usepackage{amsmath}
\usepackage{amssymb}
\usepackage{epsfig}
\usepackage{multirow}

\begin{document}

\title{Evidence for out-of-equilibrium crystal nucleation in suspensions of oppositely charged colloids}


\author{Eduardo Sanz$^{1}$\footnote{Electronic address: {\tt
e.sanz@phys.uu.nl}}, Chantal Valeriani$^{2}$, Daan Frenkel$^{2}$ and Marjolein Dijkstra$^{1}$}
\affiliation{$^{1}$Soft Condensed Matter, Utrecht University, Princetonplein 5, 3584 CC Utrecht, The Netherlands\\
$^{2}$ FOM Institute for Atomic and Molecular Physics, Kruislaan 407, 1098 SJ Amsterdam, The Netherlands}

\date{\today}

\begin{abstract}

We report a numerical study of the rate of crystal nucleation in a binary
suspension of oppositely charged colloids.
Two different crystal structures compete in the thermodynamic conditions under study.
We find that
the crystal phase that nucleates is metastable and, more surprisingly,
its nucleation free energy barrier is not the lowest one.
This implies that, during nucleation,  there is
insufficient time for sub-critical nuclei to relax to their lowest
free-energy structure. Such behavior is in direct contradiction
with the common assumption that the phase that crystallizes most
readily  is the one with the lowest free-energy barrier for
nucleation. The phenomenon that we describe should be relevant for crystallization
experiments where competing solid structures are not connected by an easy transformation.

\end{abstract}



\maketitle Liquids often must be cooled substantially below the
freezing temperature before spontaneous crystallization occurs
in the bulk.  The reason is that the system has to overcome a
free-energy barrier when moving from the metastable liquid to the
stable solid phase. When the two phases are separated by a high
free-energy barrier, spontaneous fluctuations that would result in
the formation of the stable phase are unlikely and therefore rare.
Most fluctuations will result in the formation of ephemeral
``sub-critical'' crystal nuclei that redissolve spontaneously.
Only occasionally a crystal nucleus will form that exceeds the
critical size needed for spontaneous subsequent
growth~\cite{kelton}. The crystal nucleation rate is defined as
the number of post-critical clusters  that form per unit time
in a unit volume. In classical nucleation theory (CNT), it is
assumed that sub-critical clusters are in quasi-equilibrium with
the parent phase~\cite{kelton}. This assumption is reasonable if
the time it takes to establish an equilibrium distribution of
sub-critical clusters is short compared to the time needed to
nucleate a crystal. If the nucleation rate is low, the
steady-state distribution of sub-critical clusters of size $n$ is
(nearly) proportional to $\exp(-\beta\Delta G(n))$, where $\Delta
G(n)$ is the free energy associated with the formation of a
crystalline cluster of size $n$ in the metastable liquid. The CNT
expression for the nucleation rate per unit volume is
\begin{equation}\label{eqn:rate}
R= \kappa  e^{-\Delta G_{crit}/k_{B}T}
\end{equation}
where $\kappa$ is a kinetic prefactor and $\Delta G_{crit}$ is the
height of the nucleation barrier. It was already pointed out by
Ostwald~\cite{Ostwaldrule} that often, during crystal nucleation,
a solid phase forms that is not the thermodynamically most stable
one. Stranski and Totomanow~\cite{Ostwaldrevisited} have
rationalized this observation in the language of CNT
by suggesting that the phase that nucleates is
the one separated from the parent phase by the lowest
free-energy barrier - and this need not be the most stable solid
phase. Implicit in this explanation is the assumption that the
kinetic prefactor $\kappa$ is similar for different nucleation
routes, and that hence the relative nucleation rates are
exclusively determined by the heights of the nucleation barriers.
Unfortunately, the assumptions underlying the Stranski-Totomanow
(ST) rule cannot easily be tested in experiments. Here we present
simulations where we compute independently the rate of crystal
nucleation and the height of the free energy barriers
separating a metastable liquid from two more stable solid phases.

In order to study rare events such as liquid-solid nucleation by
simulation, one has to resort to special simulation techniques,
precisely because a typical nucleation event does not occur within
the time scale of a conventional simulation. The only alternative
is to use very large system sizes~\cite{JCP_2003_119_11298} and
long simulation times~\cite{N_2002_416_00409}. But even then the
metastable system has to be prepared in a state
deeply supersaturated before spontaneous nucleation can be
observed~\cite{JPCA_1998_102_2708}.

Here, we use the the Forward-Flux-Sampling (FFS) method of Allen
et al.~\cite{PRL_2005_94_018104,JCP_2006_124_024102} to compute
the rate of crystal nucleation. This method was designed to study
rare events both {\it in} and {\it out} of equilibrium. It can be
used under conditions where brute-force simulations become
impractical. FFS has been used to calculate the rate of crystal
nucleation in molten salts~\cite{JCP_2005_122_194501} and the
nucleation rate of an Ising model in pores~\cite{PRL_2006_97_065701}.

To compute the free-energy barriers for crystal nucleation, we use
umbrella sampling~\cite{JCP_1992_96_4655}. This method has
been used before to compute the free-energy barriers for the nucleation
of crystals~\cite{PRL_1995_75_002714,JCP_2006_125_024508} and
liquids~\cite{JCP_1998_109_09901}. The umbrella-sampling approach
determines the variation of the free energy of the system with a
reaction coordinate that measures the progress of the
transformation from the liquid to the crystalline phase. It should
be noted that, whereas the nucleation {\em rate} is an observable
quantity, the height of the free-energy barrier for crystal
nucleation may depend somewhat on the choice of the reaction
coordinate.


Making use of the information obtained using both methods
we will show below that crystal
nucleation in a mixture of oppositely charged colloids is
incompatible with the ST conjecture.

\begin{figure}[!hbt]
\includegraphics[clip,height=0.13\textheight,width=0.25\textwidth,angle=-0]{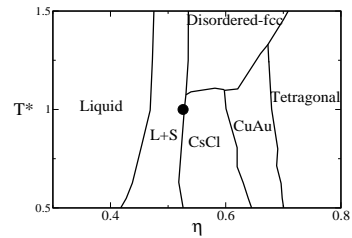}
\caption{\small Phase diagram of the system under study in the T*,
packing fraction ($\eta=\pi\sigma^3N/6V$) plane~\cite{PRL_2006_96_018303}. L+S stands for
liquid-solid coexistence. The circle indicates the state point of the
metastable liquid (T*=1, p*=15, $\eta$=0.526).The {\it disordered-fcc-fluid-CsCl} triple point temperature is $T^*=1.07$.}
\label{pd}
\end{figure}

In our simulations, we studied a 1:1 binary mixture of
monodisperse, oppositely charged colloids. The screened Coulomb
interaction between two colloids of diameter $\sigma$ and charge
$Ze$  is approximated by a Yukawa potential:
\begin{equation}
u(r)/k_BT=
\begin{cases}
\infty         & r<\sigma\\
\pm\frac{\displaystyle Z^2}{\displaystyle (1+\frac{\kappa\sigma}{2})^2}\frac{\displaystyle \lambda_B}{\displaystyle \sigma}\frac{\displaystyle e^{-\kappa(r-\sigma)}}{\displaystyle r/\sigma} & r_c > r \ge \sigma \\
0 & r \ge r_c
\end{cases}
\end{equation}
where the sign is positive for equally charged and negative for
oppositely charged colloids, $\lambda_B=e^2/\epsilon_sk_BT$ is the
Bjerrum length ($\epsilon_s$ is the dielectric constant of the
solvent) and $\kappa=\sqrt{8\pi \lambda_B \rho_{salt}}$ is the inverse
Debye screening length ($\rho_{salt}$ is the number density of
added salt). A hard core prevents colloids from overlapping. The
total energy of the system is the sum of the pair interactions. The cut-off radius, $r_c$,
is 3.5$\sigma$. We define the reduced
temperature $T^*=(1+\kappa\sigma/2)^2\sigma/Z^2\lambda_B$ as the inverse
of the contact energy, and the reduced pressure as
$p^*=pT^*\sigma^3/k_BT$. The phase diagram of this potential for
$k\sigma=6$ reproduces the solid structures that are found
experimentally in mixtures of oppositely charged colloids
\cite{PRL_2006_96_018303}. In this system, two solid phases can
coexist with the fluid. At high temperatures, the liquid phase
coexists with a substitutionally disordered face-centered cubic colloidal crystal (disordered-fcc).
At low
temperatures the stable solid at coexistence has CsCl structure,
where the charges are ordered on a bcc lattice (Fig. \ref{pd}).

This system is a suitable candidate to test the Stranski-Totomanow
conjecture as two distinct solid phases may form during crystal
nucleation. In contrast to systems that have been studied earlier~\cite{JPCM_2002_14_7667,PRL_2004_93_166105}
these two solids are not connected by an ``easy'' (e.g. martensitic)
transformation. Besides, we study nucleation close to the coexistence
temperature between both solids.

Both for the FFS calculations and for the calculation of the
free-energy barrier separating liquid and solid, we need a
reaction coordinate that measures the progress of the nucleation
process. In the present study, we use $n$, the number  of
particles in the largest solid cluster, as a reaction
coordinate~\cite{JCP_1996_104_09932}. Our reaction coordinate
distinguishes liquid from solid but is not sensitive to the
structure of the crystal lattice. In fact, under the thermodynamic
conditions used in our study ($T^*=1$, $p^*=15$), every particle
in either the disordered fcc phase or the CsCl-like solid is
identified as crystalline. On the contrary, in the metastable liquid
phase, less than five out of 1000 %
particles were identified as crystalline.

The FFS technique expresses the nucleation rate, $\phi_{l-s}$, as
the product of two factors~\cite{JCP_2003_118_7762}:
\begin{equation}
\phi_{l-s}=\phi_{l-J}P_{J-s}
\label{flux}
\end{equation}
where $\phi_{l-J}$ is the rate at which spontaneous fluctuations
lead to the formation of a small crystallite consisting of $J$
particles, whilst $P_{J-s}$ denotes the probability that such
a cluster will grow to form a bulk solid, rather than redissolve.
In what follows, we will ignore the effect of hydrodynamic
interactions and estimate $\phi_{l-J}$ using a kinetic Monte Carlo (MC)
algorithm\cite{PA_1990_166_473} with a maximum displacement of 0.01
$\sigma$ \cite{JPCM_2002_14_7667,Nature_2001_409_1020}.
In the limit of small trial displacements, the MC algorithm
approaches Brownian Dynamics, but MC has the added advantage that
we can easily work in the $NpT$ ensemble.
The probability $P_{J-s}$ is  computed as a product of probabilities:
\begin{equation}
P_{J-s}=P_{J-K}P_{K-L}...P_{N-s}
\end{equation}
where $P_{J-K}$ is the probability that a trajectory that starts
with a cluster of size $J$, will grow to
size $K$ rather than redissolve.  This probability can be
estimated by starting a number of trajectories from a cluster of
size $J$ and dividing the number of those that arrive at $K$ by
the total number of trials. The successful trajectories provide
starting configurations for the next step, namely, the calculation
of the probability that cluster $K$ will grow to size
$L$, rather than redissolve. The FFS method only works if the
dynamics of the system is not fully deterministic.  In the present
case, different kinetic MC trajectories (with maximum displacement
0.04 $\sigma$) were generated from the
same configuration by changing the seed of the random number
generator.
We stress that the ``reaction coordinate'' in the FFS
scheme is only used to measure the progress of the crystal growth --it
does not favor one crystal structure over another~\cite{PRL_2005_94_018104,JCP_2006_124_024102}.

In this work, we have studied the crystallization of the
metastable liquid phase in a system of 1000 particles at $T^*=1$
and $p^*$=15 ($p^*/p^*_{\rm coex}\approx 1.7$). The packing
fraction of the liquid at the coexistence pressure, $p_{\rm
coex}^*=8.8\pm0.1$, is $\eta=\pi\sigma^3N/6V = (0.471\pm0.005)$. At $T^*=1$ and $p^*=15$,
the packing fraction of the metastable liquid is
$(0.526\pm0.005)$. 

\begin{figure}[!hbt]
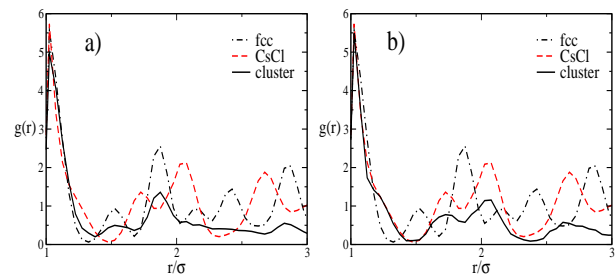

\includegraphics[clip,height=0.15\textheight,width=0.22\textwidth,angle=-0]{Fig2A.eps}
\includegraphics[clip,height=0.15\textheight,width=0.22\textwidth,angle=-0]{Fig2B.eps}
\caption{\small Radial distribution function of an 80 particles cluster obtained in a typical FFS path
at T*=1 and p*=15
compared with the radial distribution function of the solids  CsCl and disordered-fcc. (a) without swap moves, (b) with swap moves.
The structure of the clusters varies with the way in which the configurational space is sampled.
The unnormalised radial distribution functions of the clusters have been multiplied by $20$ in order to compare with the
``bulk'' radial distribution functions.}
\label{FFSclust}
\end{figure}

At $p^*=15$, no spontaneous nucleation
is observed even after $3\cdot10^6$ MC cycles (a cycle consists of
a trial move per particle and a volume move). However, the
nucleation rate can be computed at $p^*=15$ using the FFS method
(\ref{flux}). We find that the probability that a crystalline
cluster of 5 particles will continue to solidify is
$10^{-28\pm2}$. The rate at which spontaneous fluctuations in the
metastable liquid result in the formation of  crystalline clusters
of 5 particles is equal to $10^{-4\pm1}D_0\sigma^{-5}$ (where
$D_0$ denotes the diffusion coefficient at infinite dilution).
Therefore, the nucleation rate is estimated to be
$10^{-32\pm3}D_0\sigma^{-5}$. We observe the growth of a rather
compact solid cluster in the metastable liquid. When the
crystalline cluster has reached a size of $n=120\pm 15$, it has a
50\% probability of redissolving: this is our operational
definition of the critical nucleus size. To identify the crystal
structure of the solid nucleus, we analyse the radial distribution
function of the particles that belong to the cluster. This
provides a convenient way to distinguish disordered fcc and CsCl
structures. Figure \ref{FFSclust} a  shows the comparison of
$g(r)$ for a cluster of 80 particles with that of the bulk solid
phases. From the figure, it is evident that the arrangement of the
particles in the growing solid cluster is fcc-like even though the
stable solid phase at $T^*=1$ and $p^*=15$ is CsCl.

The formation of crystal nuclei of a metastable solid phase can be
interpreted as a manifestation of the Ostwald step rule. According
to the conjecture of Stranski and
Totomanow\cite{Ostwaldrevisited}, the free energy barrier for the
formation of  disordered fcc should then be lower than the one for
the formation of a CsCl cluster. The ST conjecture relies on
the assumption that sub-critical nuclei are in quasi-equilibrium.
However, as we will show, this turns out not to be the case.
We can test this by
repeating the FFS scheme with a different kinetic MC scheme that
includes an additional trial move: the swap of positive and
negative particles. If the system is already in equilibrium, the
introduction of additional MC moves will not change the structure of
the sub-critical nuclei.

When we performed  FFS simulations including 20\%  swap moves we
observe the formation of charge-ordered clusters with a CsCl
structure (Fig.\ref{FFSclust} b). Not only the structure of
the sub-critical nuclei has changed but also the size of the
critical nucleus: it now contains $65\pm15$. Moreover, the
probability that a solid cluster of 5 particles will form a bulk
crystal has increased to 10$^{-15\pm1}$.
The fact that the
pathway for crystal nucleation can be altered by artificially
improving the sampling of configurational space indicates that local
equilibrium is not established during the natural nucleation
dynamics (no swaps). This observation is in direct contradiction to the key
assumption underlying  the Stranski-Totomanow conjecture. Our
simulations suggest that the time it takes a cluster to grow from
a small size to the critical size is too short to allow for
efficient sampling of the accessible configurational space -- as a
result, it gets kinetically trapped in a metastable structure. We
expect that such behavior will be common when there is no ``easy''
kinetic route (e.g. martensitic transformation) from the
metastable to the stable crystal phase.

At higher pressures ($p^*=18$), where the probability of forming 
a post-critical nucleus is higher and
no FFS is needed to observe the transition, the same phenomenology
is reproduced.  In a kinetic MC simulation without swap moves a
substitutionally disordered fcc lattice is formed. The same is
obtained in a Brownian Dynamics simulation. In contrast, when swap
moves are included, the liquid transforms into a substitutionally
ordered lattice.

%

\begin{figure}[!hbt]
\includegraphics[clip,height=0.2\textheight,width=0.45\textwidth,angle=-0]{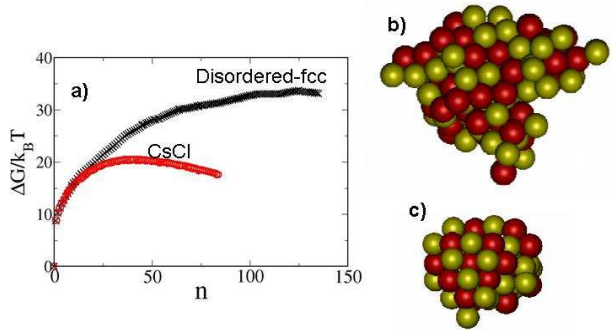}
\caption{\small (a) : Free energy barriers
calculated with umbrella sampling at T*=1 and p*=15
in  a system of 8000 particles. The CsCl clusters have lower free
   energy, but unless unphysical MC moves are used in the sampling,
    the system remains kinetically trapped in a disordered-fcc route
     of higher free energy.
     (b): Snapshot of a typical critical cluster with
     disordered-fcc structure and (c):CsCl structure. Note that the
     disordered critical cluster is bigger than the ordered one.}
\label{saltitos}
\end{figure}

Interestingly, as we can
selectively prepare fcc or CsCl nuclei by changing our kinetic MC
scheme, we can now separately compute with umbrella sampling~\cite{JCP_1992_96_4655}
the free energies of these
two different types of clusters.
In umbrella-sampling simulations, the minimum free energy path is sampled along
a given reaction coordinate.
Nevertheless, when the calculation is carried out without swap moves, the growth
of CsCl clusters is dramatically slowed
down, instead,  fcc clusters are formed and persist
for long time in the system.
If the simulation is run long enough, the structure of the
clusters changes into CsCl, suggesting that CsCl clusters
have indeed lower free energy.
By including swap
moves in the umbrella sampling scheme, clusters grow directly in their lowest free-energy state
(CsCl), yielding a different free-energy barrier.
Fig.\ref{saltitos} shows the barriers for both types of
calculations. As can be seen from the figure, the free-energy
barrier for the nucleation of fcc clusters is higher than that for
CsCl clusters.
Snapshots of two typical critical clusters (fcc and CsCl) are also shown
in Fig.~\ref{saltitos}.

This observation has direct implication for the interpretation of
experiments~\cite{PRL_2005_95_128302,N_2005_437_7056}. The
dynamics of real binary crystals of charged colloids is best
described by the kinetic MC scheme (i.e. without unphysical swap
moves). Hence we should expect that in charged colloidal systems,
crystallization proceeds through a sequence of non-equilibrium
sub-critical crystal nuclei. In experiments on crystallization in
binary charged colloids~\cite{PRL_2006_96_018303}, both
substitutionally ordered (CuAu)  and substitutionally disordered
(fcc) crystallites have been observed. The simultaneous
observation of both phases could be explained thermodynamically if
the experimental conditions fortuitously happened to  correspond
to coexistence. The present work suggests another explanation:
non-equilibrium nucleation of the fcc phase precedes a subsequent,
slow transformation to the substitutionally ordered crystal phase.

At first sight, it might seem that the present results, although
at odds with the ST conjecture, are not incompatible with
CNT. After all, within that theory,
preferred nucleation of the crystal structure with the higher
nucleation barrier is possible if a large kinetic prefactor in
Eq.~\ref{eqn:rate}  compensates the effect of the higher
nucleation barrier. Yet, the existing versions of CNT do not
correctly describe this effect: in CNT the kinetic prefactor
describes the rate at which clusters grow due to the attachment
and detachment of single particles to a pre-existing crystallite, and the
rate of addition and removal of particles is hardly different for
fcc and CsCl clusters. What seems to happen is that small clusters
have a disordered fcc structure, but this structure cannot act as
a template for subsequent CsCl growth, whilst a structural phase
transition inside the clusters is kinetically inhibited.
A crystal cluster could change its
internal structure by a succession of particle additions and
removals, but in practice this would mean that a disordered fcc
cluster would have to redissolve almost completely before it can
form a CsCl cluster.
The ``success'' of the small sub-critical fcc clusters blocks the
subsequent formation of the more stable CsCl clusters. This
phenomenon is reminiscent of the ``self-poisoning'' of small
crystallites during the rapid growth of post-critical crystal nuclei~\cite{JACS_2004_126_13347}. 
The difference is that, in the present case, the self-poisoning
already takes place with sub-critical nuclei.

The present results imply  that, at least to predict crystal
nucleation, there are situations where it is not enough to compute
the free energy barrier that separates the parent phase from
resultant solid structures - beyond a certain cluster size, the
formation of the lowest free-energy clusters may be kinetically
inhibited. The fast growth of the clusters results in the
breakdown of the local equilibrium assumption for sub-critical
nuclei.

E. S. acknowledges useful discussions with M. Hermes and A.
Cuetos. The authors acknowledge a critical reading by J. van Meel
and R. Allen.
The work of the FOM Institute is part of the research
program of FOM and is made possible by financial support from the
Netherlands organization for Scientific Research (NWO).


\bibliographystyle{./apsrev}

\end{document}